\documentclass[final,3p,times,twocolumn]{elsarticle}

\usepackage{lineno,hyperref}
\usepackage{mathrsfs}
\usepackage{amsfonts}
\usepackage{amsmath,amssymb,bm}
\usepackage{array}
\usepackage{verbatim}
\usepackage{epsfig}
\usepackage{graphicx}
\usepackage[normalem]{ulem}
\usepackage{xcolor}
\usepackage{multirow}

\usepackage{graphicx}
\graphicspath{{./figs/}}
\usepackage{dcolumn}
\usepackage{bm}
\usepackage{slashed}
\usepackage{siunitx}
\usepackage{ulem,xpatch}
\usepackage{comment}
\usepackage{soul}
\usepackage{xcolor}
\usepackage{xfrac}
\newcommand{\diff}{{\rm d}}

\usepackage{float}
\usepackage{subfigure}
\hypersetup{colorlinks, linkcolor = [rgb]{0.051,0.427,0.612}, citecolor = [rgb]{0.051,0.427,0.612}, urlcolor = [rgb]{0.051,0.427,0.612}}

\begin{document}

\begin{frontmatter}

\title{Six-dimensional light-front Wigner distribution of hadrons}

\author[PKU]{Yingda~Han}
\author[SDU]{Tianbo~Liu}
\author[PKU,CHEP]{Bo-Qiang~Ma}

\address[PKU]{School of Physics and State Key Laboratory of Nuclear Physics and Technology, Peking University, Beijing 100871, China}
\address[SDU]{Key Laboratory of Particle Physics and Particle Irradiation (MOE), Institute of Frontier and Interdisciplinary Science, Shandong University, Qingdao, Shandong 266237, China}
\address[CHEP]{Center for High Energy Physics, Peking University, Beijing 100871, China}

\begin{abstract}
We propose a six-dimensional light-front Wigner distribution for the complete description of partonic structures of a hadron such as pion and proton, taking advantage of the recently proposed light-front variable $\tilde{z}$ by Miller and Brodsky. Quantities derived from the Wigner distribution contain the most general information of partonic structures, including also new quantities correlating longitudinal coordinate with transverse momenta or transverse coordinates, together with spins. The new Wigner distribution can be viewed as a relativistic version of the original Wigner distribution in hadron physics and an extension of widely utilized five-dimensional light-front Wigner distribution.
\end{abstract}


\end{frontmatter}


\section{Introduction}

 Unraveling the structure of hadrons in terms of quarks and gluons, the fundamental degrees of freedom 
 in quantum chromodynamics (QCD), is one of the key issues in modern particle physics. The parton model, together with QCD factorization, is proven a powerful tool in explaining high-energy hadron scatterings with parton distribution function (PDF) describing the probability density of finding a parton (quark or gluon) carrying light-front (LF) longitudinal momentum fraction $x$ of the parent hadron. For the description of observables that are also sensitive to the transverse kinematics of a parton, the concept of PDF has been generalized to transverse momentum dependent parton distributions (TMDs) and generalized parton distributions (GPDs) to include the information of transverse momentum and transverse coordinate distributions respectively, towards a multi-dimensional tomography of the nucleon, as one of the main goals of the upcoming Electron-Ion Collider~\cite{Accardi:2012qut,AbdulKhalek:2021gbh,Anderle:2021wcy}.

As has been known for a long time, the most complete information of a system is encoded in quantum phase space distributions, among which the Wigner distribution is the most widely used. It is originally introduced in non-relativistic quantum mechanics~\cite{Wigner:1932eb},
\begin{align}
    \rho({\bm b}, {\bm k}) = \int \diff^3 {\bm y}
    e^{i{\bm k}\cdot{\bm y}}
    \psi^*\Big({\bm b}-\frac{\bm y}{2}\Big)
    \psi \Big({\bm b}+\frac{\bm y}{2}\Big),
\end{align}
and the expectation value of any physical operator can be obtained from the average over the Wigner distribution~\cite{Hillery:1983ms}.
Due to the Heisenberg's uncertainty principle in quantum theories, the Wigner distribution is a quasi-distribution, describing non-positive definite density of finding the particle at average position $\bm b$ with average momentum $\bm k$.
This concept has been applied in various areas, such as the quantum information, quantum molecular dynamics, optics, nonlinear dynamics, quantum transport, and heavy ion collisions~\cite{Balazs:1983hk}, 
and was introduced to hadron physics in Refs.~\cite{Ji:2003ak,Belitsky:2003nz} as six-dimensional functions by neglecting relativistic effects.
For a proper description of the nucleon, which is a relativistic bound state of the strong interaction, the LF Wigner distribution is defined~\cite{Lorce:2011kd} as five-dimensional function of the longitudinal momentum fraction $x$, transverse position ${\bm b}_\perp$, and transverse momentum ${\bm k}_\perp$. Taking advantage of the LF dynamics~\cite{Dirac:1949cp}, or sometimes interpreted as the physics in the infinite momentum frame~\cite{Weinberg:1966jm}, the LF Wigner distribution is boost-invariant, connecting to TMDs by integrating over transverse position and to GPDs at zero skewness, $\xi=0$, by integrating over transverse momentum. Although the measurement of Wigner distributions is much more challenging than the measurements of TMDs and GPDs, several processes such as the diffractive di-jet production in deep inelastic scattering~\cite{Hatta:2016dxp}, the virtual photon-nucleus quasi-elastic scattering~\cite{Zhou:2016rnt}, and the exclusive double Drell-Yan process~\cite{Bhattacharya:2017bvs}, have been proposed in recent years, which increases our confidence in extracting the rich information encoded in Wigner distributions from future experiments.
The measurements of generalized transverse momentum dependent parton distributions (GTMDs) at nonzero skewness through these processes may also shed light on the extraction of the proposed six-dimentional Wigner distribution via Fourier transform.

Comparing with the originally Wigner distribution in non-relativistic quantum mechanics, one easily notices that the LF Wigner distribution does not include all position and momentum variables in pair, missing the longitudinal position. This seems a natural consequence of the Lorentz contraction in special relativity. However, the LF coordinate $b^-$, which contracts to zero in the infinite momentum frame, is not the right variable to describe the longitudinal position of a parton. One needs to introduce a boost-invariant variable related to $b^-$, similar to the case that we use $x$, instead of $k^+$, to describe the longitudinal momentum of the parton. Such variable is recently proposed by Miller and Brodsky~\cite{Miller:2019ysh}, $\tilde{z}=b^-P^+$, where $P^+$ is the LF momentum of the hadron. We note that this variable was first considered in the study of deeply virtual Compton scattering (DVCS) process in Refs.~\cite{Brodsky:2006in,Brodsky:2006ku}, which provide an analogy to the diffraction pattern in optics. 
The boost-invariant LF longitudinal position variable $\tilde{z}$ is merely a parameter, which is not associated with an operator. This inherits the general statement that one cannot construct a position operator in relativistic quantum field theory, because any attempt to localize a particle will involve high frequency modes whose energy is enough to create additional particles. The closest concept one may consider as a position observable is the center of inertia~\cite{Lorce:2018zpf}.
In this letter, we demonstrate that the Miller-Brodsky variable is exactly the conjugate variable to the skewness $\xi$ in GPDs or GTMDs. Hence, the LF Wigner distribution can be extended by pairing the longitudinal momentum with the Miller-Brodsky variable to provide a complete information of parton distributions inside the hadron.

\section{Extended light-front Wigner distributions.}
\label{sec:eWD}

We start from the Wigner operator, similar to the one in Ref.~\cite{Lorce:2011kd} but including the dependence on LF longitudinal coordinate,
\begin{align}
    \widehat{W}^{[\Gamma]}
    &(b^-, {\bm b}_\perp, k^+, {\bm k}_\perp)
    = \int \frac{\diff y^- \diff^2{\bm y}_\perp}{2(2\pi)^3}
e^{ik^+y^- - i{\bm k}_\perp\cdot {\bm y}_\perp}
    \nonumber\\
    &\times
    \bar{\psi}\left( b - \frac{y}{2}\right)\Gamma
    {\cal L}_{[b-y/2,b+y/2]}
    \psi\left(b + \frac{y}{2}\right)\Bigg|_{y^+=0},
    \label{eq:Woperator}
\end{align}
where $b=(0,b^-,{\bm b}_\perp)$ is the average LF position at some LF time which we set as $0$, and ${\cal L}_{[b-y/2,b+y/2]}$ is the Wilson line connecting the quark operators to ensure the color gauge invariance. $\Gamma$ is a Dirac matrix projecting the quark spin state, and at the leading twist, $\Gamma=\gamma^+$, $\gamma^+\gamma_5$, and $i\sigma^{i+}\gamma_5$, corresponds to unpolarized, longitudinally polarized, and transversely polarized quarks respectively. The condition $y^+=0$ reflects the probe at a fixed LF time. The extended LF Wigner distribution is then defined by interpolating the Wigner operator~\eqref{eq:Woperator} between initial and final nucleon states with a momentum transfer $\Delta=(\Delta^+,\Delta^-,{\bm \Delta}_\perp)$,
\begin{align}
    \rho^{[\Gamma]}&(\tilde{z},x,{\bm b}_\perp,{\bm k}_\perp,S)
    = \int \frac{\diff \xi \diff^2{\bm \Delta}_\perp}{4\pi^3}
    \nonumber\\
    &\times\Big\langle P + \frac{\Delta}{2}, S\Big|
    \widehat{W}^{[\Gamma]}(b^-,{\bm b}_\perp, k^+, {\bm k}_\perp)
    \Big|P - \frac{\Delta}{2}, S\Big\rangle,
    \label{eq:Wdistribution}
\end{align}
where $\xi=-\Delta^+/(2P^+)$ is the skewness variable representing the longitudinal momentum transfer in a physical process, $P^+$ is the average LF momentum of the nucleon, $x=k^+/P^+$ is the average LF momentum fraction carried by the quark, $\tilde{z}=b^-P^+$ is the Miller-Brodsky variable reflecting the longitudinal position, and $S$ is the spin state of the nucleon. Keeping quark momentum fraction non-negative, the integral over $\xi$ runs from $-x$ to $x$, which corresponds to the so-called DGLAP region~\cite{Diehl:2003ny}, {\it i.e.}, the transferred momentum $\Delta^+$ is not enough to produce a quark pair each carrying average momentum fraction $x$.
Then the function defined in~\eqref{eq:Wdistribution} represents the quantum phase space distribution of a single quark inside the nucleon.

To see the relation between
$\rho^{[\Gamma]}(\tilde{z},x,{\bm b}_\perp,{\bm k}_\perp,S)$ and the widely used LF Wigner distribution, one can insert translation operations between the Wigner operator and the nucleon state and integrate over $\tilde{z}$,
\begin{align}
    &\int \diff \tilde{z}\,
    \rho^{[\Gamma]}(\tilde{z},x,{\bm b}_\perp,{\bm k}_\perp, S)
    = -\int \diff b^- \int \frac{\diff {\Delta^+} \diff^2{\bm \Delta}_\perp}{(2\pi)^3}
    \nonumber\\
    &\times e^{ib^-\Delta^+}
    \Big\langle P + \frac{\Delta}{2}, S\Big|
    \widehat{W}^{[\Gamma]}(0,{\bm b}_\perp, k^+, {\bm k}_\perp)
    \Big|P - \frac{\Delta}{2}, S\Big\rangle
    \nonumber\\
    &=\int \frac{\diff^2{\bm \Delta}_\perp}{(2\pi)^2}
    \Big\langle P + \frac{{\bm \Delta}_\perp}{2}, S\Big|
    \widehat{W}^{[\Gamma]}(0,{\bm b}_\perp, k^+, {\bm k}_\perp)
    \Big|P - \frac{{\bm \Delta}_\perp}{2}, S\Big\rangle,
\end{align}
which is just the five-dimensional LF Wigner distribution as introduced in Ref.~\cite{Lorce:2011kd}. Therefore, the extend Wigner distribution in Eq.~\eqref{eq:Wdistribution} is a generalization of the well-known five-dimensional Wigner distribution without losing any information, but including also the longitudinal position distribution and its correlations with $x$, ${\bm b}_\perp$, ${\bm k}_\perp$, and spins.

Similar to the five-dimensional version, the extended LF Wigner distributions have a direct connection with the GTMDs,
\begin{align}
    &\rho^{[\Gamma]}(\tilde{z}, x, {\bm b}_{\perp}, {\bm k}_{\perp}, S)
    \nonumber\\
    &=\int\frac{\diff\xi\diff^{2}{\bm \Delta}_{\perp}}{4 \pi^{3}}
    e^{-2i\xi\tilde{z}-i{\bm b}_{\perp}\cdot{\bm \Delta}_{\perp}}
    W^{[\Gamma]}\big(\xi, x, {\bm \Delta}_{\perp}, {\bm k}_{\perp}, S\big),
    \label{eq:W-GTMD}
\end{align}
where $W^{[\Gamma]}(\xi, x,{\bm \Delta}_{\perp}, {\bm k}_{\perp}, S)$ are the GTMDs introduced in Refs.~\cite{Meissner:2008ay,Meissner:2009ww},
\begin{align}
& W^{[\Gamma]}\big(\xi, x, {\bm \Delta}_{\perp}, {\bm k}_{\perp}, S\big)
= \int \frac{\diff y^- \diff^2{\bm y}_\perp}{2(2\pi)^3}
    e^{ixP^+y^- - i{\bm k}_\perp\cdot {\bm y}_\perp}
    \nonumber\\
    &\times\Big\langle P + \frac{\Delta}{2}, S\Big|
    \bar{\psi}\left(- \frac{y}{2}\right)\Gamma
    {\cal L}_{[-y/2,+y/2]}
    \psi\left(+\frac{y}{2}\right)\Big|P - \frac{\Delta}{2}, S\Big\rangle\Big|_{y^+=0}.
    \label{eq:GTMD}
\end{align}
While GTMDs are complex functions, one can easily find that the extended Wigner distributions are real functions.
From~\eqref{eq:W-GTMD}, one can observe that the Miller-Brodsky variable $\tilde{z}$ is conjugate to the skewness $\xi$. Integrating out the transverse momentum ${\bm k}_\perp$ and the transverse position ${\bm b}_\perp$ of~\eqref{eq:Wdistribution}, one can define the longitudinal LF Wigner distribution,
\begin{align}
    \rho^{[\Gamma]}(\tilde{z},x)
    = \int \diff^2{\bm k}_\perp \diff^2{\bm b}_\perp
    \rho^{[\Gamma]}&(\tilde{z},x,{\bm b}_\perp,{\bm k}_\perp,S),
    \label{eq:W-zx}
\end{align}
which is a two-dimensional quasi-distribution function encoding partonic distribution information along the longitudinal direction. As we will show later, this function is not positive definite. The probability interpretation clearly fails for negative values. It is due to the non-classicality in quantum theory, or in other words, it reflects how well a classical description works for a parton at certain kinematics.
The non-negativity is equivalent to non-contextuality, originating from the description of quantum phenomena by so-called hidden variable models in which each observable has a pre-determined value merely revealed by the act of measurement. As explained in Refs.~\cite{Spekkens_2008,Okay2020homotopicalapproach,blass2015negative}, it is impossible in a quantum theory. Alternatively, one can view the Wigner distribution as the kernel of the density matrix~\cite{Hiley_2016}.

For a multi-dimensional tomography of the hadron, one needs distribution functions including transverse kinematics. Apart from TMDs and GPDs (at $\xi=0$) that can be obtained via the integrals of the five-dimensional LF Wigner distributions, we can also define three-dimensional parton distribution functions by integrating the extended LF Wigner distributions over ${\bm k}_\perp$ and $x$,
\begin{align}
    \rho^{[\Gamma]}(\tilde{z}, {\bm b}_\perp, S)
    = \int \diff x \diff^2{\bm k}_\perp \rho^{[\Gamma]}(\tilde{z}, x, {\bm b}_\perp, {\bm k}_\perp, S),
    \label{eq:3d-zbt}
\end{align}
as position-space counterparts to TMDs. One can also define longitudinal position-transverse momentum joint three-dimensional parton distribution functions by integrating over ${\bm b}_\perp$ and $x$,
\begin{align}
    \rho^{[\Gamma]}(\tilde{z}, {\bm k}_\perp, S)
    = \int \diff x \diff^2{\bm b}_\perp \rho^{[\Gamma]}(\tilde{z}, x, {\bm b}_\perp, {\bm k}_\perp, S),
    \label{eq:3d-zkt}
\end{align}
as counterparts to GPDs which describe the parton distribution in a joint three-dimensional space of longitudinal momentum and transverse position. In addition, one can learn the correlation of the longitudinal position with transverse momentum, transverse position, and spins directly from the extended Wigner distributions. Unlike TMDs and GPDs at zero skewness, the three-dimensional distributions in Eqs.~\eqref{eq:3d-zbt} and~\eqref{eq:3d-zkt} do not have probability interpretations. The on-shell condition of the initial and final state hadrons, $(P - \Delta/2)^2 = (P + \Delta/2)^2 = M^2$, requires $P\cdot \Delta = P^+ \Delta^- + P^- \Delta^+ = 0$. To obtain the distribution in $\tilde{z}$, one needs the amplitude at nonzero skewness $\xi = -\Delta^+/(2P^+) \neq 0$, which also leads to nonzero $\Delta^-$ and thus $x^+$ dependence.

\section{Modeling the extended LF Wigner distribution.}

To further our understanding, we perform calculation of the extended LF Wigner distribution of the pion and the proton in the spectator model, which is proven successful in qualitatively describing many physical quantities, such as form factors, unpolarized and polarized quark PDFs, and TMDs. Applying the LF Fock-state expansion to the hadron state,
\begin{align}
    \left|P,S\right\rangle =&
    \sum_{n} \int [\diff x \diff^2 {\bm k}_\perp]
    \Psi\left(x_i,{\bm k}_{i\perp},\lambda_{i};n\right)
    \nonumber\\
    &\times
    \left|n:x_{i} P^{+},x_{i}{\bm P}_{\perp}+{\bm k}_{i\perp},\lambda_{i}\right\rangle,
    \label{eq:Fock}
\end{align}
where $n$ runs over all possible Fock states, $x_i$, ${\bm k}_{i\perp}$ and $\lambda_i$ represent the LF momentum fraction, intrinsic transverse momentum and helicity of the constituent within the Fock state, $\Psi\left(x_i,{\bm k}_{i\perp},\lambda_{i};n\right)$ is the LF wave function (LFWF), and the integral measure is
\begin{align}
    [\diff x \diff^2 {\bm k}_\perp]
    &= \left(16 \pi^{3}\right)
    \left(\prod_{i\in n}\frac{\diff x_{i} \diff^{2} {\bm k}_{i\perp}}{2 \sqrt{x_{i}}(2 \pi)^{3}}\right)
    \nonumber\\
    &\times
     \delta\left(1-\sum_{i\in n} x_{i}\right) \delta^{(2)}\left(\sum_{i\in n} {\bm k}_{i\perp}\right),
\end{align}
one can in principle calculate the extend LF Wigner distributions following the definition in~\eqref{eq:Wdistribution} if LFWFs are provided.

Since solving LFWFs is not the scope of this work, we adopt the spectator model LFWFs, which have been utilized in many phenomenological studies. For the pion, we take the valence state LFWFs~\cite{Xiao:2003wf,Ma:2018ysi},
\begin{subequations}
\begin{align}
    \Psi\left(x, {\bm k}_{\perp}, \uparrow, \downarrow\right)
    &= -\Psi\left(x, {\bm k}_{\perp}, \downarrow, \uparrow\right)
    = \frac{m}{\sqrt{2{\cal M}_T^2}} \varphi_{\pi}, \\
    \Psi\left(x, {\bm k}_{\perp}, \uparrow, \uparrow\right)
    &= \Psi^*\left(x, {\bm k}_{\perp}, \downarrow, \downarrow\right)
    = -\frac{k_{1}-i k_{2}}{\sqrt{2{\cal M}_T^2}} \varphi_{\pi},
    \label{eq:pionLFWFs}
\end{align}
\end{subequations}
where ${\cal M}_T^2=m^2+{\bm k}_\perp^2$, the arrows represent the helicities of quark and the spectator antiquark, and $\varphi_\pi$ is the spin-independent wave function which we choose the Brodsky-Huang-Lepage prescription~\cite{Brodsky:1980vj},
    $\varphi_{\pi} = A_\pi \exp \left\{-{\cal M}_T^2/[8\beta^{2}x(1-x)]\right\}$
with $A_\pi$ as the normalization factor. The parameters are $m=0.2\,\rm GeV$ and $\beta=0.41\,\rm GeV$ are chosen according to Ref.~\cite{Ma:2018ysi}.

For the proton, we consider the valence component as a quark-diquark configuration, including both scalar and axial-vector diquarks which serve as a spectator and absorb part of high Fock-state contributions into the effective masses. The explicit expressions of LFWFs are chosen the same as those in Ref.~\cite{Liu:2015eqa}, where the momentum space wave functions are taken as real functions sharing the same form of the Brodsky-Huang-Lepage prescription.

For simplicity, we consider the unpolarized extended Wigner distribution, which for the pion case is defined as $\rho_{\rm UU}^{\pi}(\tilde{z},x,{\bm b}_\perp,{\bm k}_\perp) = \rho^{[\gamma^+]}(\tilde{z},x,{\bm b}_\perp,{\bm k}_\perp)$, and for the proton case is defined as the average over spin states,
\begin{align}
    \rho_{\rm UU}(\tilde{z},x,{\bm b}_\perp,{\bm k}_\perp)
    &= \frac{1}{2} \Big[
    \rho^{[\gamma^+]}(\tilde{z},x,{\bm b}_\perp,{\bm k}_\perp, S)
    \nonumber\\
    &+ \rho^{[\gamma^+]}(\tilde{z},x,{\bm b}_\perp,{\bm k}_\perp, -S)
    \Big].
    \label{eq:rho-UU}
\end{align}
Substituting the Fock expansion \eqref{eq:Fock} into \eqref{eq:Wdistribution}, one can derive the overlap representation of the extended LF Wigner distribution of the pion as
\begin{align}
    &\rho_{\rm UU}(\tilde{z},x,{\bm b}_\perp,{\bm k}_\perp)
    =
    \sum_{\lambda_q,\lambda_{\bar q}}
    \int\frac{\diff\xi\diff^{2}{\bm \Delta}_{\perp}}{4\pi^{3}}
    e^{-2i\xi\tilde{z}-i{\bm b}_{\perp}\cdot{\bm \Delta}_{\perp}}
    \nonumber\\
    &~\times
    \frac{1}{16\pi^3}
    \Psi^*\left(x^{\rm out}, {\bm k}_\perp^{\rm out}, \lambda_q, \lambda_{\bar q}\right)
    \Psi\left(x^{\rm in}, {\bm k}_\perp^{\rm in}, \lambda_q, \lambda_{\bar q}\right),
    \label{eq:rho-UU-wf}
\end{align}
where $x^{\rm out}=(x-\xi)/(1-\xi)$ and $x^{\rm in}=(x+\xi)/(1+\xi)$ are longitudinal momentum fractions carried by the struck quark in the final and initial states.
The transverse momenta
${\bm k}_{\perp}^{\rm out}={\bm k}_{\perp}+(1-x){\bm \Delta}_\perp/[2(1-\xi)]$
and
${\bm k}_{\perp}^{\rm in}={\bm k}_{\perp}-(1-x){\bm \Delta}_\perp/[2(1+\xi)]$
in the LFWFs are intrinsic transverse momenta of the struck quark with respect to the final and initial state hadrons.
Their relations to ${\bm k}_\perp$ are derived by substituting the Fock state expansion~\eqref{eq:Fock} into Eq.~\eqref{eq:Wdistribution}.
In the frame where the final state hadron has transverse momentum ${\bm \Delta}_\perp/2$ and correspondingly the initial state hadron has transverse momentum $-{\bm \Delta}_\perp/2$, the transverse momenta of the struck quark can be evaluated from the intrinsic transverse momenta via 
\begin{align}
{\bm p}_\perp^{\rm out} &= {\bm k}_\perp^{\rm out} + x^{\rm out} \frac{{\bm \Delta}_\perp}{2} = {\bm k}_\perp + \frac{{\bm \Delta}_\perp}{2},\\
{\bm p}_\perp^{\rm in} &= {\bm k}_\perp^{\rm in} - x^{\rm in} \frac{{\bm \Delta}_\perp}{2} = {\bm k}_\perp - \frac{{\bm \Delta}_\perp}{2},
\end{align}
with the difference equaling to the transferred transverse momentum ${\bm p}_\perp^{\rm out} - {\bm p}_\perp^{\rm in} = {\bm \Delta}_\perp$. The $\lambda_{q,\bar{q}}$ are helicities of the quark and the spectator antiquark. Similar expressions can be obtained for the proton case, where one needs to sum over the spectator types and spin states.
Corresponding unpolarized longitudinal LF Wigner distribution $\rho_{\rm UU}(\tilde{z},{x})$, three-dimensional position distribution function $\rho_{\rm UU}(\tilde{z}, {\bm b}_\perp)$, and longitudinal position-transverse momentum joint three-dimensional distribution function $\rho_{\rm UU}(\tilde{z},{\bm k}_\perp)$ are defined as the integrals of~\eqref{eq:rho-UU} following~\eqref{eq:W-zx},~\eqref{eq:3d-zbt}, and~~\eqref{eq:3d-zkt} respectively.

\begin{figure}[htbp]
\centering
\includegraphics[width=3.3cm]{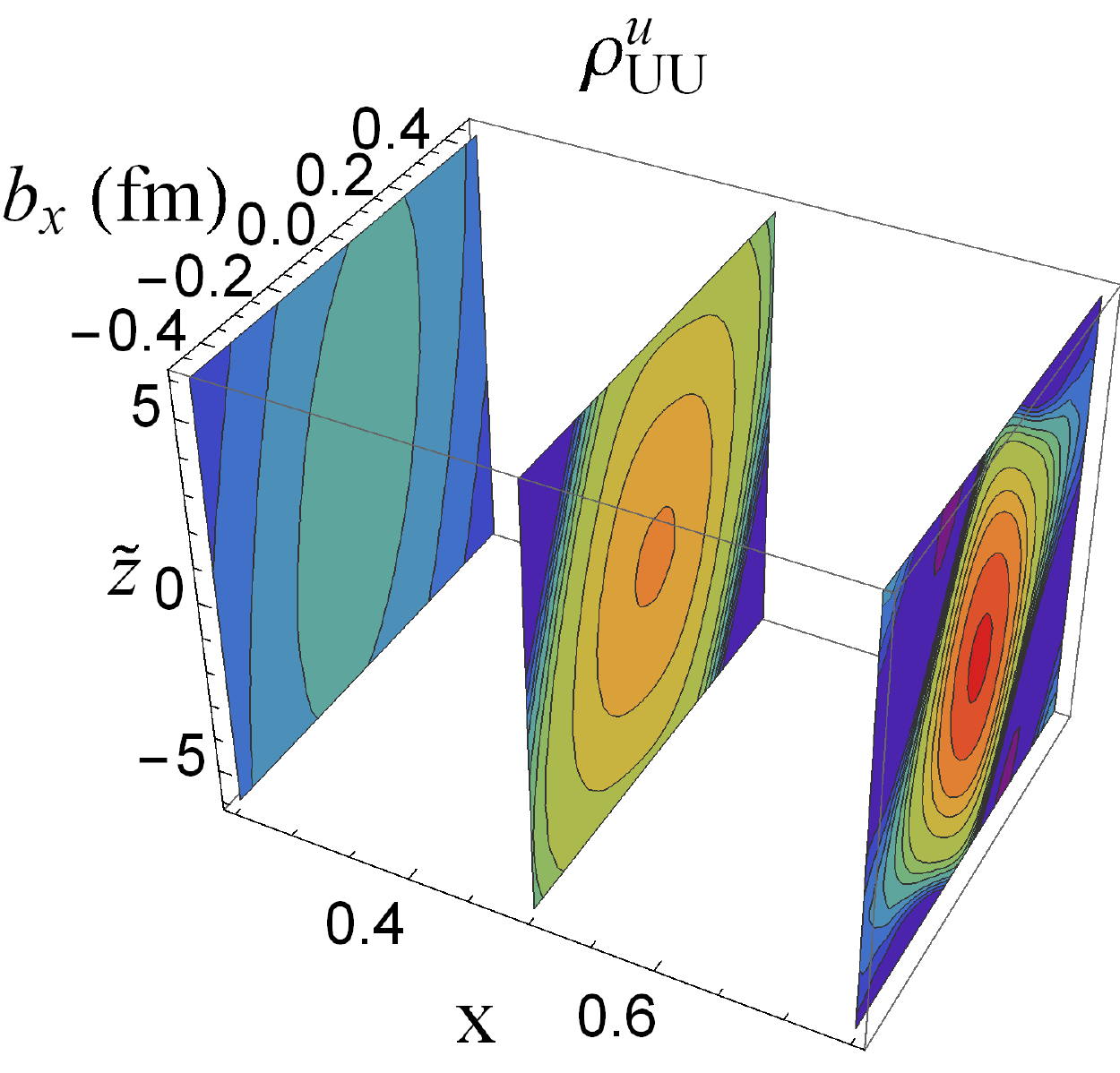}
\includegraphics[width=3.3cm]{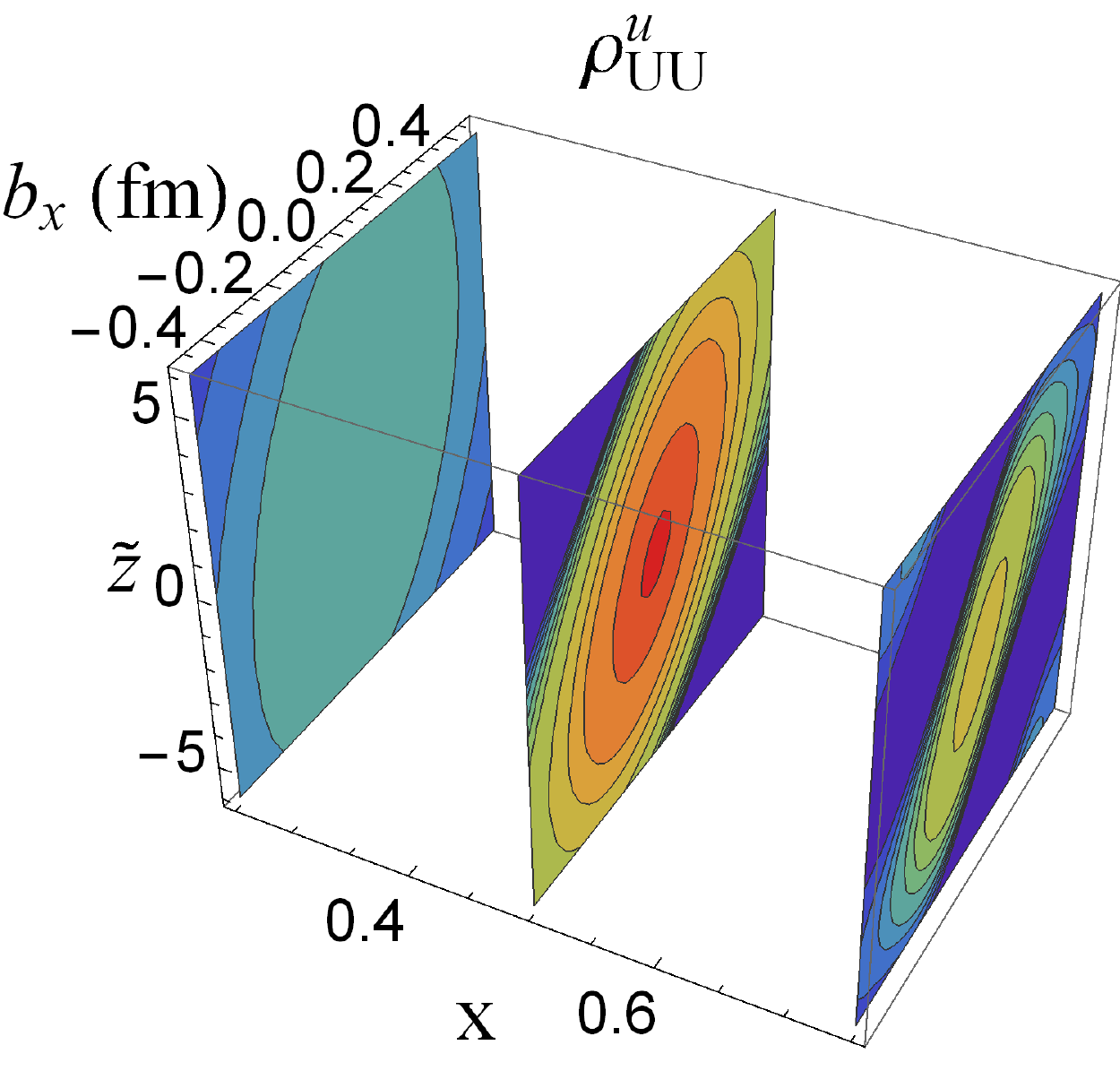}
\includegraphics[width=0.9cm]{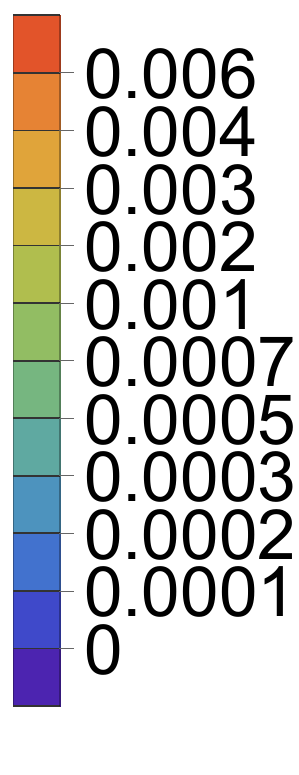}\\
\includegraphics[width=3.3cm]{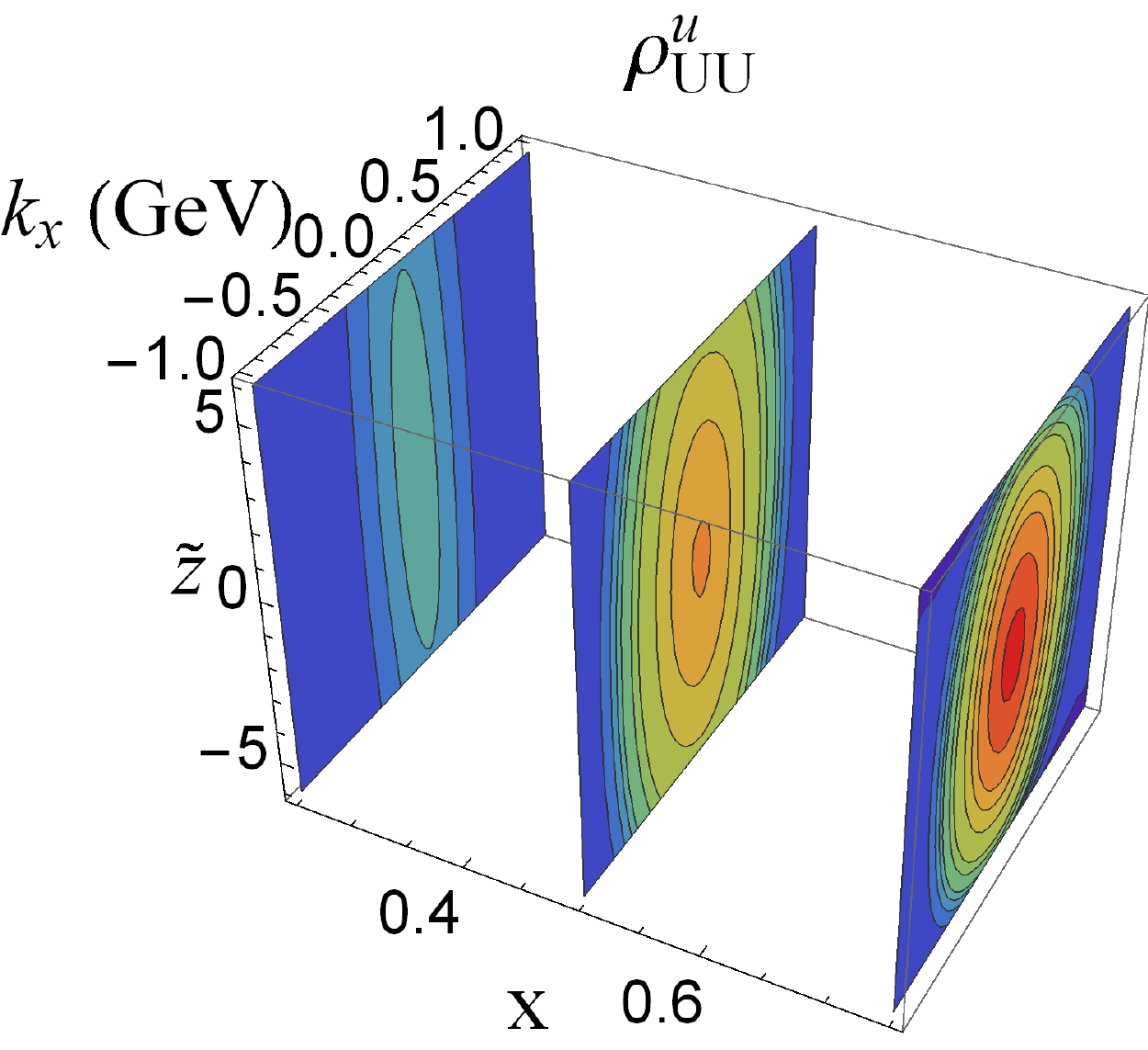}
\includegraphics[width=3.3cm]{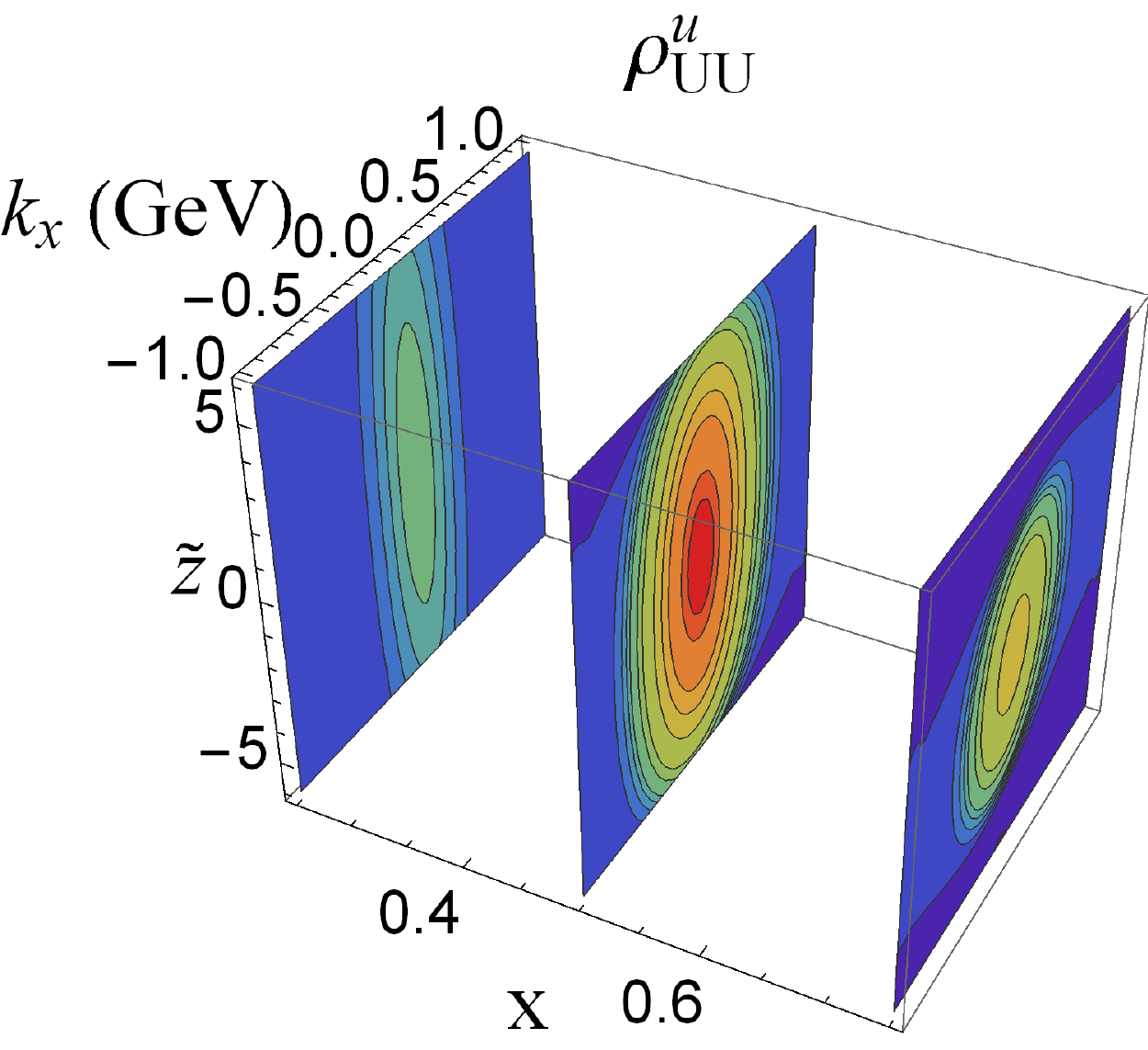}
\includegraphics[width=0.9cm]{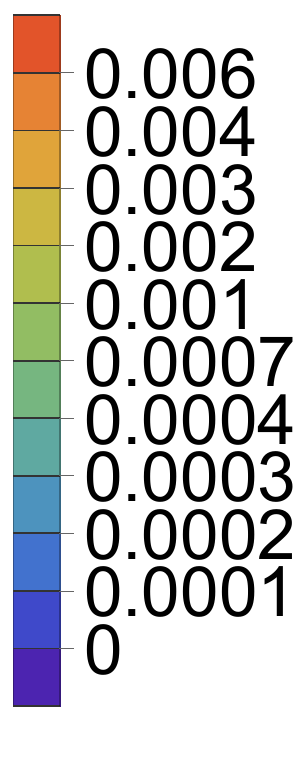}
\caption{Six-dimensional unpolarized LF Wigner distribution 
$\rho_{\rm UU}(\tilde{z},x, {\bm b}_\perp, {\bm k}_\perp)$ of the pion (left) and the proton (right).
Upper : the distribution in $\tilde{z}$-$b_x$ plane with fixed $\bm{k}_{\perp}=0.6\,{\rm GeV}\,{\bm e}_x$ (${\bm e}_x$ is the unit vector in $x$ direction) and $b_{y}=0.4\,\rm GeV^{-1}$. 
Lower : the distribution in $\tilde{z}$-$k_x$ plane with fixed $\bm{b}_{\perp}=0.4\,{\rm GeV^{-1}}\,{\bm e}_x$ and  $k_{y}=0.6\,\rm  GeV$. }
\label{fig:W-6d}
\end{figure}

\begin{figure}[hbp]
\centering
\includegraphics[width=\columnwidth]{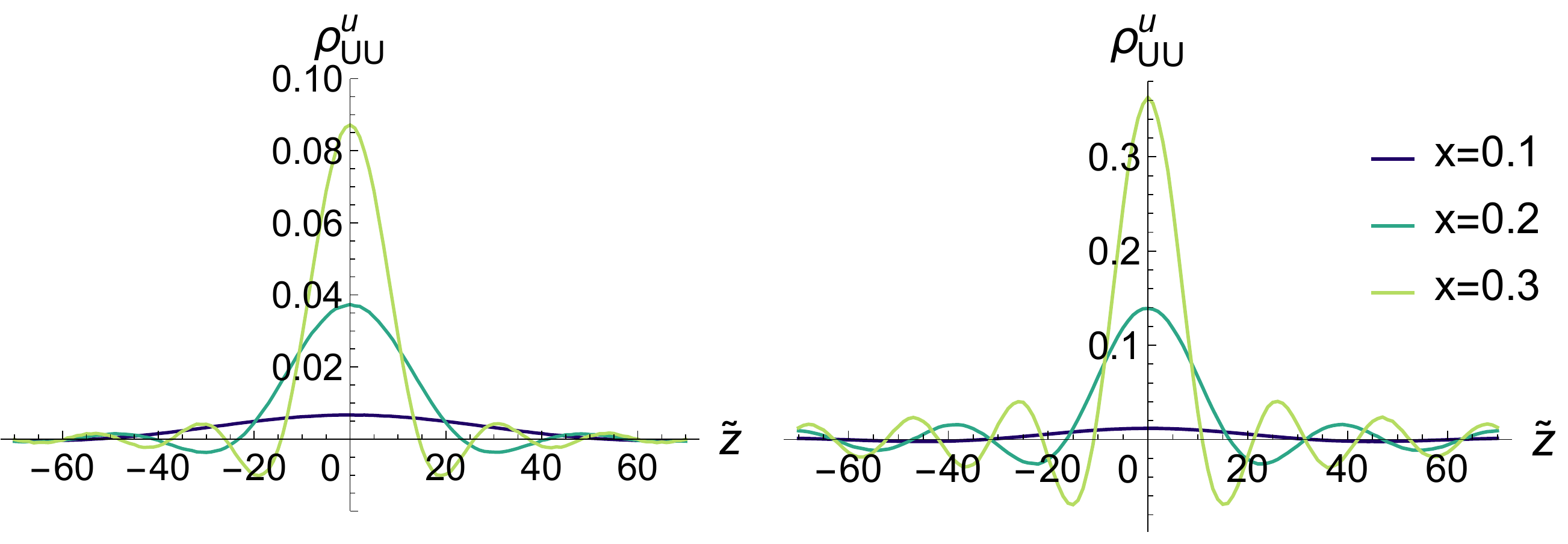}
\caption{Unpolarized longitudinal LF Wigner distributions $\rho_{\rm UU}(\tilde{z},x)$ of the pion (left) and the proton (right).}
\label{fig:W-zx}
\end{figure}

The numerical results of unpolarized extend LF Wigner distribution $\rho_{\rm UU}(\tilde{z}, x, {\bm b}_\perp, {\bm k}_\perp)$ of the pion and the proton are shown in Fig.~\ref{fig:W-6d}. 
To see the correlation of longitudinal position with transverse momentum and transverse position, the function is plotted at some fixed values of ${\bm k}_\perp$ or ${\bm b}_\perp$ as specified in the caption. 
The unpolarized distribution is centro-symmetry in the $\tilde{z}$-$\bm b_{\perp}$ and $\tilde{z}$-$\bm k_{\perp}$ subspaces, which can be derived from~\eqref{eq:rho-UU-wf}. One can also observe in Fig.~\ref{fig:W-6d} that the distribution around the central region increases at large $x$, which can be roughly interpreted with the picture that a parton carrying larger momentum is less likely to appear far from the center. Since the range of $\xi$ becomes wider with increasing $x$, the distribution in its Fourier conjugate $\tilde{z}$ is more centralized as a typical feature of the Fourier transform.
We should note that the value of the distribution around the $\tilde{z}$ center is a balance between the decrease of $\tilde{z}$-integrated distribution ({\it i.e.} the PDF) at large $x$ and the centralization of the distribution in $\tilde{z}$.
This feature is more clearly shown in Fig.~\ref{fig:W-zx}, the unpolarized longitudinal LF Wigner distribution $\rho_{\rm UU}(\tilde{z},x)$. With increasing $x$ values, the distribution becomes more and more centralized in $\tilde{z}$. 
At longitudinal positions away from the center (the classical limit), the Wigner distribution is negative in some regions, which is interpreted as nonclassical behavior of the quantum system.
In addition, the oscillating behavior of the distribution in $\tilde{z}$ can be viewed as an analog to the diffraction pattern in optics. Such analogy was first found in the study of the DVCS amplitude in the longitudinal distance~\cite{Brodsky:2006in, Brodsky:2006ku}. The finite range of the skewness $\xi$, from $-x$ to $x$, is like the width of the slit in the single slit experiment, which leads to the oscillating behavior in the Fourier transform. As can be observed from the curves at different $x$ values in Fig.~\ref{fig:W-zx}, the distribution oscillates in $\tilde{z}$ and achieves its first local minimal point at smaller $\tilde{z}$ value with increasing $x$ value.

\begin{figure}[t]
\centering
\includegraphics[width=3.8cm]{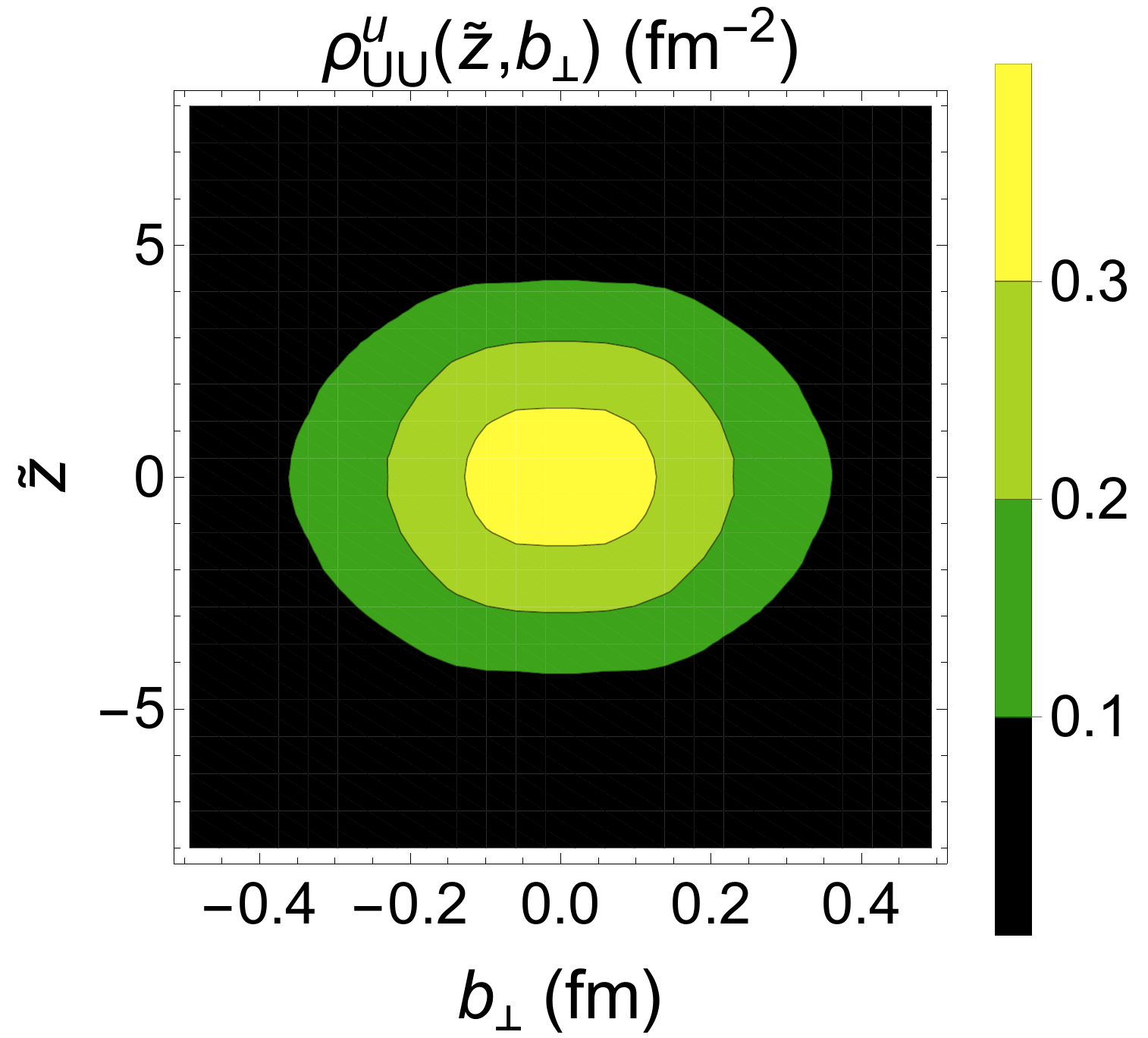}
\includegraphics[width=3.8cm]{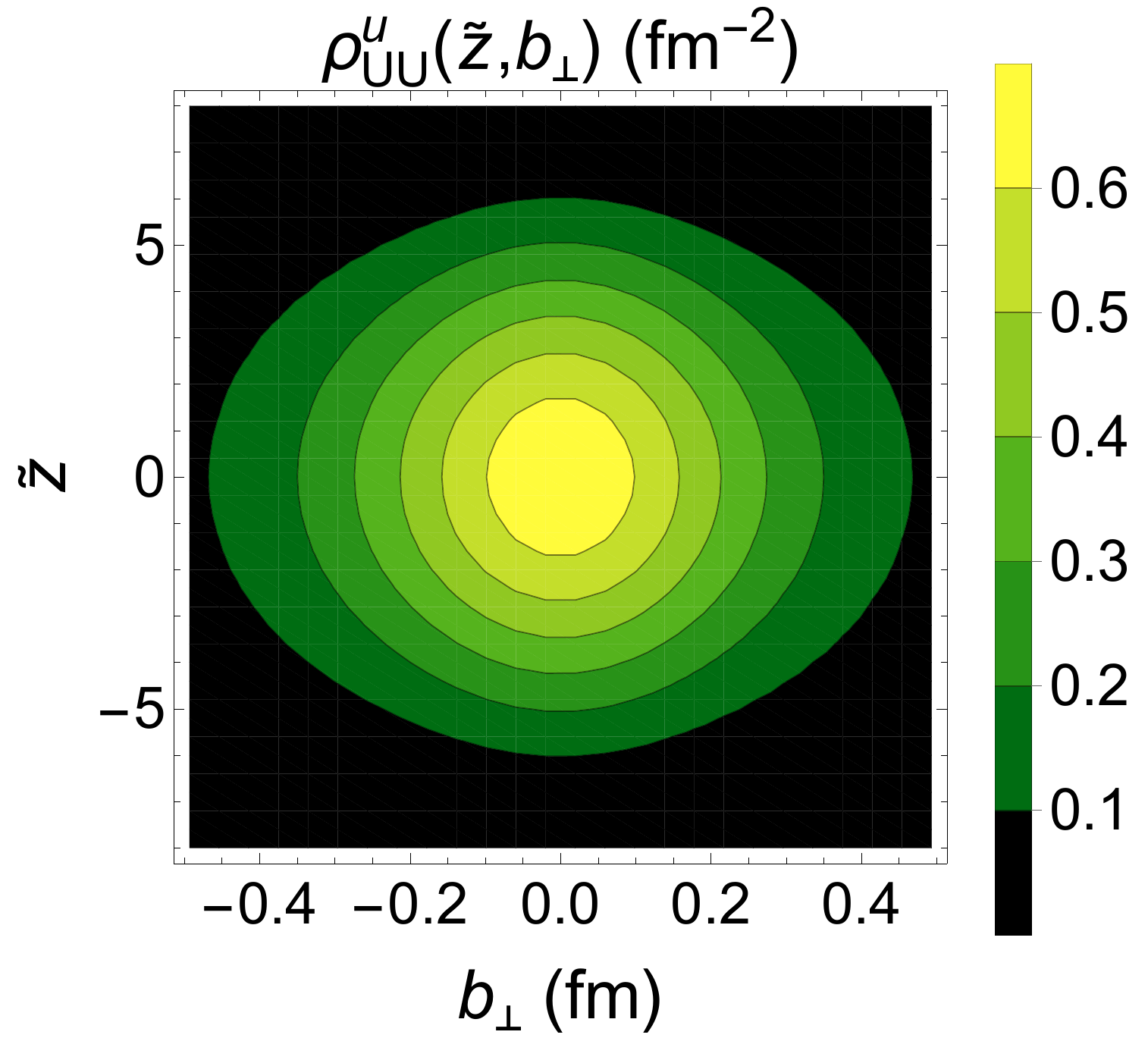}
\caption{Three-dimensional parton distribution function $\rho_{\rm UU}(\tilde{z}, {\bm b}_{\perp})$ of the pion (left) and the proton (right).}
\label{fig:3d-zbt}
\end{figure}

\begin{figure}[t]
\centering
\includegraphics[width=3.8cm]{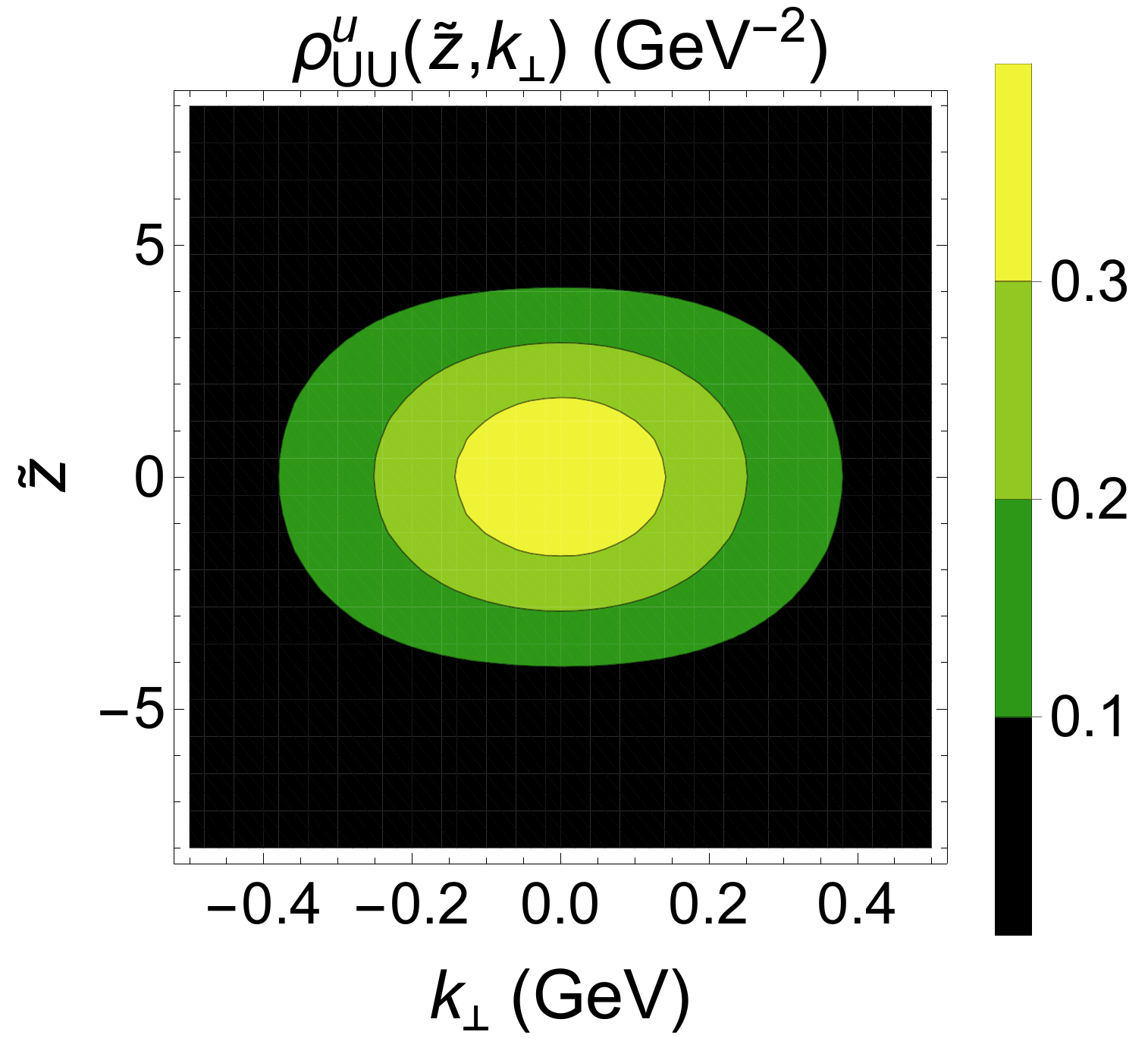}
\includegraphics[width=3.8cm]{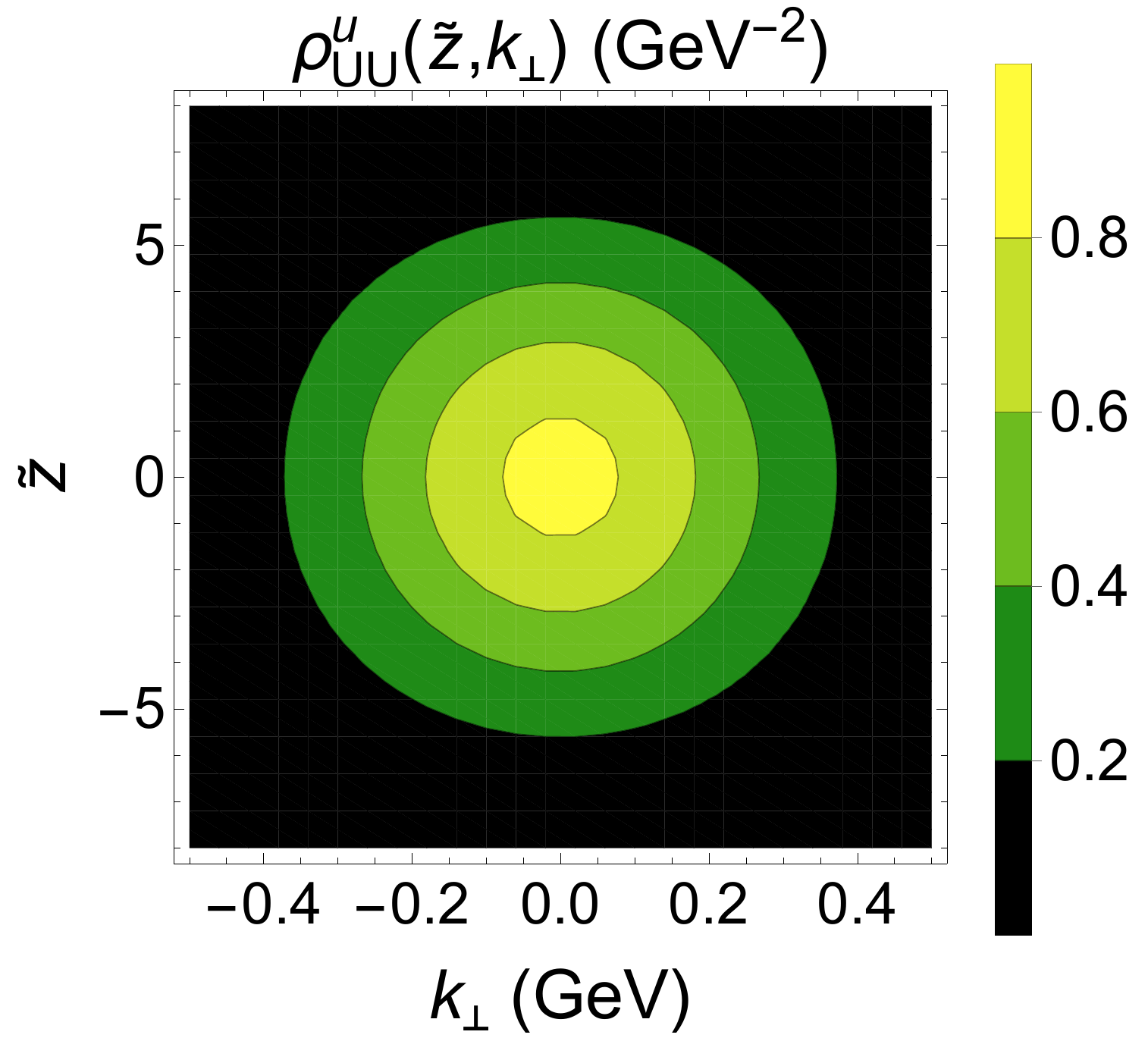}
\caption{Three-dimensional parton distribution function $\rho_{\rm UU}(\tilde{z}, {\bm k}_{\perp})$ of the pion (left) and the proton (right).}
\label{fig:3d-zkt}
\end{figure}

The unpolarized three-dimensional position distribution functions $\rho_{\rm UU}(\tilde{z},{\bm b}_\perp)$ of the pion and the proton are shown in Fig.~\ref{fig:3d-zbt}, and the unpolarized longitudinal position-transverse momentum joint three-dimensional distribution functions $\rho_{\rm UU}(\tilde{z},{\bm k}_\perp)$ of the pion and the proton are shown in Fig.~\ref{fig:3d-zkt}. These functions are symmetric in  the  longitudinal  position,  transverse  position and transverse momentum, and have a positive peak at the center. They represent the correlations of the longitudinal position with transverse momentum and transverse position respectively. As mentioned in section~\ref{sec:eWD}, $\rho_{\rm UU}(\tilde{z},{\bm b}_\perp)$ and $\rho_{\rm UU}(\tilde{z},{\bm k}_\perp)$ do not have probability interpretations. In Figs. 3 and 4, these distributions are only plotted in the region $|\tilde{z}| \leq 8$, where the values are positive. If going to large $\tilde{z}$, one will find nonpositive values and oscillating behavior. Such long and oscillating tail crossing zero of the distribution in $\tilde{z}$ was also found in the form factor calculation in the $(1+1)$-dimensional $\phi^3$ model~\cite{Choi:2021xld}.

\section{Summary and outlook.}

We have proposed extended LF Wigner distributions for the description of hadron structures. With all position and momentum variables appear in pair, they are six-dimensional functions, which contain the most complete information of parton distributions in the hadron. The boost-invariant LF variable $\tilde{z}$, referred to as the Miller-Brodsky variable, is conjugate to the skewness $\xi$ in GPDs and GTMDs, allowing us to define relativistic Wigner distributions beyond fixing $\xi=0$.
As an extension of the widely used five-dimensional LF Wigner distribution, one can learn richer information of partonic structures, particularly the longitudinal position distribution and its correlation with transverse momentum, transverse position, and spins. From these extended LF Wigner distributions, we are able to define new three-dimensional parton functions that encodes the correlation of longitudinal position with transverse momentum or transverse position, in addition to the TMDs and GPDs that are being actively studied.

For an illustration of these new quantities, we take the unpolarized case as an example with spectator model calculation. As expected, non-positivity is observed in longitudinal LF Wigner distribution, but from the quasi-distribution point of view a general feature that quarks at larger $x$ have more centralized distribution in the longitudinal position $\tilde{z}$, is obtained in consistent with intuitions. 

When the spin degree of freedom is taken into account, one is able to learn much richer partonic structures of the nucleon, and to have a more profound understanding of the strong interaction in confining phase. Considering a transversely polarized nucleon, the longitudinal position-transverse momentum joint three-dimensional function will shed light on the transverse orbital angular momentum, which is an indispensable part of the nucleon spin structure. 
The measurements of GTMDs at nonzero skewness via the diffractive di-jet production in deep inelastic scattering~\cite{Hatta:2016dxp}, the virtual photon-nucleus quasi-elastic scattering~\cite{Zhou:2016rnt}, and the exclusive double Drell-Yan process~\cite{Bhattacharya:2017bvs} will provide valuable information, and it is necessary to measure GTMDs at nonzero skewness for the extended LF Wigner distributions.
We leave these topics to more careful future studies.
\\

\noindent
\section*{Acknowledgements}
This work is supported in part by National Natural Science Foundation of China under Grant No. 12075003 and No. 12175117.


\begin{thebibliography}{}

\bibitem{Accardi:2012qut}
A.~Accardi {\it et al.},
Electron-Ion Collider: The next QCD frontier, Eur. Phys. J. A 52 (9) (2016) 268, \href{https://doi.org/10.1140/epja/i2016-16268-9}{https://doi.org/10.1140/epja/i2016-16268-9}.

\bibitem{AbdulKhalek:2021gbh}
R.~Abdul Khalek {\it et al.},
Science Requirements and Detector Concepts for the Electron-Ion Collider: EIC Yellow Report, \tt arXiv:2103.05419 [physics.ins-det], \href{https://arxiv.org/abs/2103.05419}{https://arxiv.org/abs/2103.05419}.

\bibitem{Anderle:2021wcy}
D.~P.~Anderle {\it et al.},
Electron-ion collider in China, Front. Phys. 16 (6) (2021) 64701, \href{https://doi.org/10.1007/s11467-021-1062-0}{https://doi.org/10.1007/s11467-021-1062-0}.

\bibitem{Wigner:1932eb}
E.~P.~Wigner,
On the Quantum Correction For Thermodynamic Equilibrium, Phys. Rev. 40 (1932) 749-760, \href{https://doi.org/10.1103/PhysRev.40.749}{https://doi.org/10.1103/PhysRev.40.749}.

\bibitem{Hillery:1983ms}
M.~Hillery, R.~F.~O'Connell, M.~O.~Scully, and E.~P.~Wigner,
Distribution functions in physics: Fundamentals., Phys. Rep. 106 (1984) 121-167, \href{https://doi.org/10.1016/0370-1573(84)90160-1}{https://doi.org/10.1016/0370-1573(84)90160-1}.

\bibitem{Balazs:1983hk}
See {\it e.g.}, 
N.~L.~Balazs, and B.~K.~Jennings,
Wigner's Function and Other Distribution Functions in Mock Phase Spaces, 104 (1984) 347, \href{https://doi.org/10.1016/0370-1573(84)90151-0}{https://doi.org/10.1016/0370-1573(84)90151-0};

M.~Hillery, R.~F.~O'Connell, M.~O.~Scully, and E.~P.~Wigner,
Distribution functions in physics: Fundamentals, Phys. Rep. 106 (1984) 121, \href{https://doi.org/10.1016/0370-1573(84)90160-1}{https://doi.org/10.1016/0370-1573(84)90160-1};

K.~Vogel and H.~Risken,
Determination of quasiprobability distributions in terms of probability distributions for the rotated quadrature phase, Phys. Rev. A 40 (1989) 2847, \href{https://doi.org/10.1103/PhysRevA.40.2847}{https://doi.org/10.1103/PhysRevA.40.2847};

D.~T.~Smithey, M.~Beck, M.~G.~Raymer, and A.~Faridani,
Measurement of the Wigner distribution and the density matrix of a light mode using optical homodyne tomography: Application to squeezed states and the vacuum, Phys. Rev. Lett. 70 (1993) 1244, \href{https://doi.org/10.1103/PhysRevLett.70.1244}{https://doi.org/10.1103/PhysRevLett.70.1244};

%
G. Breitenbach, S. Schiller, and J. Mlynek,
Measurement of the quantum states of squeezed light, Nature 387 (1997) 471, \href{https://doi.org/10.1038/387471a0}{https://doi.org/10.1038/387471a0};

K.~Banaszek, C.~Radzewicz, K.~Wodkiewicz and J.~S.~Krasinski,
Direct measurement of the Wigner function by photon counting, Phys. Rev. A 60 (1999) 674, \href{https://doi.org/10.1103/PhysRevA.60.674}{https://doi.org/10.1103/PhysRevA.60.674};

U.~W.~Heinz,
Kinetic theory for plasmas with non-Abelian interactions, Phys. Rev. Lett. 51 (1983) 351, \href{https://doi.org/10.1103/PhysRevLett.51.351}{https://doi.org/10.1103/PhysRevLett.51.351};

H.~T.~Elze, M.~Gyulassy, and D.~Vasak,
Transport Equations for the {QCD} Quark Wigner Operator, Nucl. Phys. B 276 (1986) 706-728, \href{https://doi.org/10.1016/0550-3213(86)90072-6}{https://doi.org/10.1016/0550-3213(86)90072-6};

D.~Vasak, M.~Gyulassy, and H.~T.~Elze,
Quantum Transport Theory for Abelian Plasmas, Ann. Phys. 173 (1987) 462-492, \href{https://doi.org/10.1016/0003-4916(87)90169-2}{https://doi.org/10.1016/0003-4916(87)90169-2};

P.~Zhuang and U.~W.~Heinz,
Relativistic quantum transport theory for electrodynamics, Ann. Phys. 245 (1996) 311-338, \href{https://doi.org/10.1006/aphy.1996.0011}{https://doi.org/10.1006/aphy.1996.0011};

J.~H.~Gao, Z.~T.~Liang, S.~Pu, Q.~Wang, and X.~N.~Wang,
Chiral Anomaly and Local Polarization Effect from Quantum Kinetic Approach, Phys. Rev. Lett. 109 (2012) 232301, \href{https://doi.org/10.1103/PhysRevLett.109.232301}{https://doi.org/10.1103/PhysRevLett.109.232301};

J.~h.~Gao and Q.~Wang,
Magnetic moment, vorticity-spin coupling and parity-odd conductivity of chiral fermions in 4-dimensional Wigner functions, Phys. Lett. B 749 (2015) 542-546, \href{https://doi.org/10.1016/j.physletb.2015.08.058}{https://doi.org/10.1016/j.physletb.2015.08.058};

S.~Z.~Yang, J.~H.~Gao, Z.~T.~Liang, and Q.~Wang,
Second-order charge currents and stress tensor in a chiral system, Phys. Rev. D 102 (11) (2020) 116024, \href{https://doi.org/10.1103/PhysRevD.102.116024}{https://doi.org/10.1103/PhysRevD.102.116024}.

\bibitem{Ji:2003ak}
X.~d.~Ji,
Viewing the Proton through ``Color" Filters, Phys. Rev. Lett. 91 (2003) 062001, \href{https://doi.org/10.1103/PhysRevLett.91.062001}{https://doi.org/10.1103/PhysRevLett.91.062001}.

\bibitem{Belitsky:2003nz}
A.~V.~Belitsky, X.~d.~Ji, and F.~Yuan,
Quark imaging in the proton via quantum phase-space distributions, Phys. Rev. D 69 (2004) 074014, \href{https://doi.org/10.1103/PhysRevD.69.074014}{https://doi.org/10.1103/PhysRevD.69.074014}.

\bibitem{Lorce:2011kd}
C.~Lorc\'e and B.~Pasquini,
Quark Wigner distributions and orbital angular momentum, Phys. Rev. D 84 (2011) 014015, \href{https://doi.org/10.1103/PhysRevD.84.014015}{https://doi.org/10.1103/PhysRevD.84.014015}.

\bibitem{Dirac:1949cp}
P.~A.~M.~Dirac,
Forms of Relativistic Dynamics, Rev. Mod. Phys. 21 (1949) 392-399, \href{https://doi.org/10.1103/RevModPhys.21.392}{https://doi.org/10.1103/RevModPhys.21.392}.

\bibitem{Weinberg:1966jm}
S.~Weinberg,
Dynamics at Infinite Momentum, Phys. Rev. 150 (1966) 1313-1318, \href{https://doi.org/10.1103/PhysRev.150.1313}{https://doi.org/10.1103/PhysRev.150.1313}.

\bibitem{Hatta:2016dxp}
{Y.~Hatta, B.-W.~Xiao, and F.~Yuan,
Probing the Small-$x$ Gluon Tomography in Correlated Hard Diffractive Dijet Production in Deep Inelastic Scattering, Phys. Rev. Lett. 116 (20) (2016) 202301}, \href{https://doi.org/10.1103/PhysRevLett.116.202301}{https://doi.org/10.1103/PhysRevLett.116.202301}.

\bibitem{Zhou:2016rnt}
J.~Zhou,
Elliptic gluon generalized transverse-momentum-dependent distribution inside a large nucleus, Phys. Rev. D 94 (11) (2016) 114017, \href{https://doi.org/10.1103/PhysRevD.94.114017}{https://doi.org/10.1103/PhysRevD.94.114017}.

\bibitem{Bhattacharya:2017bvs}
S.~Bhattacharya, A.~Metz, and J.~Zhou,
Generalized TMDs and the exclusive double Drell\textendash{}Yan process, Phys. Lett. B 771 (2017) 396-400, \href{https://doi.org/10.1016/j.physletb.2017.05.081}{https://doi.org/10.1016/j.physletb.2017.05.081}
[Erratum: B \textbf{810} (2020) 135866], \href{https://doi.org/10.1016/j.physletb.2020.135866}{https://doi.org/10.1016/j.physletb.2020.135866}.

\bibitem{Miller:2019ysh}
G.~A.~Miller and S.~J.~Brodsky,
Frame-independent spatial coordinate $\tilde{z}$: Implications for light-front wave functions, deep inelastic scattering, light-front holography, and lattice QCD calculations, Phys. Rev. C 102 (2) (2020) 022201, \href{https://doi.org/10.1103/PhysRevC.102.022201}{https://doi.org/10.1103/PhysRevC.102.022201}.

\bibitem{Brodsky:2006in}
S.~J.~Brodsky, D.~Chakrabarti, A.~Harindranath, A.~Mukherjee and J.~P.~Vary,
Hadron optics: Diffraction patterns in deeply virtual Compton scattering, Phys. Lett. B \textbf{641} (2006) 440-446, \href{https://doi.org/10.1016/j.physletb.2006.08.061}{https://doi.org/10.1016/j.physletb.2006.08.061}.

\bibitem{Brodsky:2006ku}
S.~J.~Brodsky, D.~Chakrabarti, A.~Harindranath, A.~Mukherjee and J.~P.~Vary,
Hadron optics in three-dimensional invariant coordinate space from deeply virtual compton scattering,
Phys. Rev. D \textbf{75} (2007) 014003, \href{https://doi.org/10.1103/PhysRevD.75.014003}{https://doi.org/10.1103/PhysRevD.75.014003}.

\bibitem{Lorce:2018zpf}
C.~Lorc\'e,
The relativistic center of mass in field theory with spin,
Eur. Phys. J. C \textbf{78} (2018) no.9, 785, \href{https://doi.org/10.1140/epjc/s10052-018-6249-3}{https://doi.org/10.1140/epjc/s10052-018-6249-3}.

\bibitem{Diehl:2003ny}
See {\it e.g.},
 M.~Diehl,
Generalized parton distributions, Phys. Rep. 388 (2003) 41-277, \href{https://doi.org/10.1016/j.physrep.2003.08.002}{https://doi.org/10.1016/j.physrep.2003.08.002}.

\bibitem{Meissner:2008ay}
S.~Mei{\ss}ner, A.~Metz, M.~Schlegel, and K.~Goeke,
Generalized parton correlation functions for a spin-0 hadron, J. High Energy Phys. 08 (2008) 038, \href{https://doi.org/10.1088/1126-6708/2008/08/038}{https://doi.org/10.1088/1126-6708/2008/08/038}.

\bibitem{Meissner:2009ww}
S.~Mei{\ss}ner, A.~Metz, and M.~Schlegel,
Generalized parton correlation functions for a spin-1/2 hadron, J. High Energy Phys. 08 (2009) 056, \href{https://doi.org/10.1088/1126-6708/2009/08/056}{https://doi.org/10.1088/1126-6708/2009/08/056}.


\bibitem{Spekkens_2008}
R. W. Spekkens,
Negativity and Contextuality are Equivalent Notions of Nonclassicality, Phys. Rev. Lett. 101 (2008) (2) 020401, \href{https://doi.org/10.1103/physrevlett.101.020401}{https://doi.org/10.1103/physrevlett.101.020401}.

\bibitem{Okay2020homotopicalapproach}
O. Cihan and R. Robert, Homotopical approach to quantum contextuality, Quantum 4 (2020) 217
, \href{https://doi.org/10.22331/q-2020-01-05-217}{https://doi.org/10.22331/q-2020-01-05-217}.

\bibitem{blass2015negative}
A. Blass and Y. Gurevich,
Negative probability, \tt arXiv:1502.00666 [quant-ph], \href{https://arxiv.org/abs/1502.00666}{https://arxiv.org/abs/1502.00666}.

\bibitem{Hiley_2016}
B. J. Hiley,
The Algebraic Way,  
in I.~Licata and G.~'t~Hooft,
\href{https://doi.org/10.1142/p1045}{\it Beyond Peaceful Coexistence}, 
Imperial College Press, 2016, pp. 1-25, \href{https://doi.org/10.1142/9781783268320_0002}{https://doi.org/10.1142/9781783268320$\_$0002}.



\bibitem{Xiao:2003wf}
B.-W.~Xiao and B.-Q.~Ma,
Pion-photon and photon-pion transition form factors in the light-cone formalism, Phys. Rev. D \textbf{68} (2003) 034020, \href{https://doi.org/10.1103/PhysRevD.68.034020}{https://doi.org/10.1103/PhysRevD.68.034020}.

\bibitem{Ma:2018ysi}
Z.-L.~Ma and Z.~Lu,
Quark Wigner distribution of the pion meson in light-cone quark model, Phys. Rev. D \textbf{98} (5) (2018) 054024, \href{https://doi.org/10.1103/PhysRevD.98.054024}{https://doi.org/10.1103/PhysRevD.98.054024}.

\bibitem{Brodsky:1980vj}
S.~J.~Brodsky, T.~Huang, and G.~P.~Lepage,
The Hadronic Wave Function in Quantum Chromodynamics, \tt SLAC-PUB-2540,\href{https://www.slac.stanford.edu/pubs/slacpubs/2500/slac-pub-2540.pdf}{https://www.slac.stanford.edu/pubs
/slacpubs/2500/slac-pub-2540.pdf}; 
T. Huang, B.-Q. Ma, Q.-X. Shen, Analysis of the pion wave function in light cone formalism, Phys. Rev. D 49 (1994) 1490-1499, \href{https://doi.org/10.1103/PhysRevD.49.1490}{https://doi.org/10.1103/
PhysRevD.49.1490}.


\bibitem{Liu:2015eqa}
T.~Liu and B.-Q.~Ma,
Quark Wigner distributions in a light-cone spectator model, Phys. Rev. D \textbf{91} (2015) 034019, \href{https://doi.org/10.1103/PhysRevD.91.034019}{https://doi.org/10.1103/PhysRevD.91.034019}.

\bibitem{Choi:2021xld}
Y.~Choi, H.~M.~Choi, C.~R.~Ji and Y.~Oh,
Light-front dynamic analysis of the longitudinal charge density using the solvable scalar field model in (1+1) dimensions,
Phys. Rev. D \textbf{103} (2021) 076002, \href{https://doi.org/10.1103/PhysRevD.103.076002}{https://doi.org/10.1103/PhysRevD.103.076002}.




\end{thebibliography}
\end{document}